\documentclass{article}
\usepackage{spconf}

\usepackage[final]{graphicx}
\usepackage{times}
\usepackage{amsfonts}
\usepackage{epsfig}
\usepackage{amsbsy,amssymb}
\usepackage{amsmath}
\usepackage{citesort}
\usepackage{inputenc}

\newcommand{\beq}{\begin{equation}}
\newcommand{\eeq}{\end{equation}}
\newcommand{\ben}{\begin{enumerate}}
\newcommand{\een}{\end{enumerate}}
\newcommand{\bit}{\begin{itemize}}
\newcommand{\eit}{\end{itemize}}

\title{2D Linear Precoded OFDM for future Mobile Digital Video Broadcasting}

\name{Oudomsack Pierre Pasquero, Matthieu Crussi\`ere, Youssef Nasser, Jean-Fran\c{c}ois H\'elard}
\address{Institute of Electronics and Telecommunications in Rennes \\
				INSA Rennes, 20, avenue des Buttes de Coesmes, 35043 Rennes, France \\
				E-mail: oudomsack.pasquero@ens.insa-rennes.fr, \{first name. last name\}@insa-rennes.fr } 
				
\begin{document}

\maketitle

\begin{abstract}
In this paper, we propose a novel channel estimation technique based on 2D spread pilots. The merits of this technique are its simplicity, its flexibility regarding the transmission scenarios, and the spectral efficiency gain obtained compared to the classical pilot based estimation schemes used in DVB standards. We derive the analytical expression of the mean square error of the estimator and show it is a function of the autocorrelation of the channel in both time and frequency domains. The performance evaluated over a realistic channel model shows the efficiency of this technique which turns out to be a promising channel estimation for the future mobile video broadcasting systems.
\end{abstract}

\section{INTRODUCTION}
Orthogonal frequency division multiplexing (OFDM) has been widely adopted in most of the digital video broadcasting standards as DVB-T \cite{DVBT}, DMB-T, ISDB-T. This success is due to its robustness to frequency selective fading and to the simplicity of the equalization function of the receiver. Indeed, by implementing inverse fast Fourier transform (IFFT) at the transmitter and FFT at the receiver, OFDM splits the single channel into multiple, parallel intersymbol interference (ISI) free subchannels. Therefore, each subchannel, also called subcarrier, can be easily equalized by only one coefficient.

To equalize the signal, the receiver needs to estimate the channel frequency response for each subcarrier. In the DVB-T standard, some subcarriers are used as pilots and interpolating filtering techniques are applied to obtain the channel response for any subcarrier. Nevertheless, these pilots reduce the spectral efficiency of the system. To limit this problem, we propose to add a two dimensions (2D) linear precoding (LP) function before the OFDM modulation. The basic idea is to dedicate one of the precoding sequences to transmit a so-called spread pilot \cite{Cariou-IEE07} that will be used for the channel estimation. The merits of this channel estimation technique are not only due to the resource conservation possibility, but also to the flexibility offered by the adjustable time and frequency spreading lengths. In addition, note that the precoding component can be exploited to reduce the peak-to-average ratio (PAPR) of the OFDM system \cite{Nobilet}, or to perform the frequency synchronisation.

The contribution of this article is twofold. First, a general framework is proposed to describe the 2D precoding technique used for channel estimation. Secondly, exploiting some properties of random matrix and free probability theories \cite{Tse-trans-InfoTheory} \cite{Debbah-trans-InfoTheory}, an analytical study of the proposed estimation method is presented.

The article is organized as follows. In section 2, we present the principles of 2D LP OFDM, and detail the channel estimation technique using the spread pilots. In section 3, we analyse the theoretical performance of this channel estimation by developing the analytical expression of its mean square error (MSE). Then, simulation results in terms of MSE and bit error rate (BER) are presented and discussed in section 4. Concluding remarks are given in section 5.

\section{SYSTEM DESCRIPTION}

\subsection{2D LP OFDM}
\begin{figure} [t]
	\begin{center}
		\includegraphics[width=1 \linewidth]{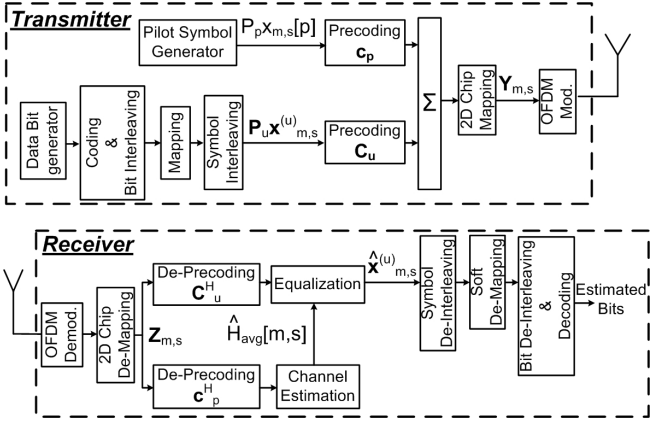}
		\caption{2D LP OFDM transmitter and receiver based on spread pilot channel estimation technique}
		\label{TX_RX}
	\end{center}
\end{figure}

Fig. \ref{TX_RX} exhibits the proposed 2D LP OFDM system exploiting the spread pilot channel estimation technique. First of all, data bits are encoded, interleaved and converted to complex symbols $x_{m,s}\left[i\right]$. These data symbols are assumed to have zero mean and unit variance. They are interleaved before being precoded by a Walsh-Hadamard (WH) sequence $\textbf{c}_{i}$ of $L$ chips, with $0 \leq i \leq L=2^{n}$ and $n \in \mathbb{N}$. The chips obtained are mapped over a subset of $L=L_{t}.L_{f}$ subcarriers, with $L_{t}$ and $L_{f}$ the time and frequency spreading factors respectively. 

\begin{figure} [t]
	\begin{center}
		\includegraphics[width=0.7 \linewidth]{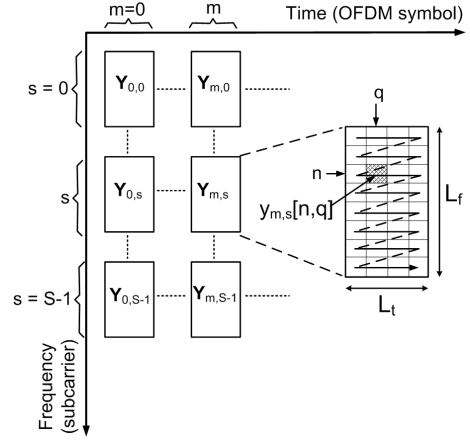}
		\caption{2D chip mapping scheme}
		\label{ChipMapping}
	\end{center}
\end{figure}

The first $L_{t}$ chips are allocated in the time direction. The next blocks of $L_{t}$ chips are allocated identically on adjacent subcarriers as illustrated in Fig. \ref{ChipMapping}. Therefore, the 2D chip mapping follows a zigzag in time. Let us define a frame as a set of $L_{t}$ adjacent OFDM symbols, and a sub-band as a set of $L_{f}$ adjacent subcarriers. In order to distinguish the different subsets of subcarriers, we define $m$ and $s$ the indexes referring to the frame and the sub-band respectively, with $0\leq s \leq S-1$. Given these notations, each chip $y_{m,s}\left[n,q\right]$ represents the complex symbol transmitted on the $n$th subcarrier during the $q$th OFDM symbol of the subset of subcarriers $\left[m,s\right]$, with $0\leq n \leq L_{f}-1$ and $0\leq q\leq L_{t}-1$. Hence, the transmitted signal on a subset of subcarriers $\left[m,s\right]$ writes:
\begin{equation}
\textbf{Y}_{m,s} = \textbf{C} \textbf{P} \textbf{x}_{m,s}
\end{equation}
where \small $\textbf{x}_{m,s} = \left[ x_{m,s}\left[0\right] \dots x_{m,s}\left[i\right] \dots x_{m,s}\left[L-1\right] \right]^{T}$ is the \small $\left[L\times1\right]$ \normalsize complex symbol vector, \small $\textbf{P} = diag \left\{ \sqrt{P_{0}} \dots \sqrt{P_{i}} \dots \sqrt{P_{L-1}} \right\}$ \normalsize is a \small $\left[L\times L\right]$ \normalsize diagonal matrix where $P_{i}$ is the power assigned to symbol $x_{m,s}\left[i\right]$, and $\textbf{C} = \left[\textbf{c}_{0} \dots \textbf{c}_{i} \dots \textbf{c}_{L-1}\right]$ is the WH precoding matrix whose $i$th column corresponds to $i$th precoding sequence \small $\textbf{c}_{i} = [ c_{i}\left[0,0\right] \dots c_{i}\left[n,q\right] \dots c_{i}[L_{f}-1,L_{t}-1] ]^{T}$. \normalsize We assume normalized precoding sequences, i.e. $c_{i}\left[n,q\right] = \pm\frac{1}{\sqrt{L}}$. Since the 2D chip mapping applied follows a zigzag in time, $c_{i}\left[n,q\right]$ is the $(n\times L_{t}+q)$th chip of the $i$th precoding sequence $\textbf{c}_{i}$.  

\subsection{Spread pilot channel estimation principles}
Inspired by pilot embedded techniques \cite{Chin-Globecom01}, channel estimation based on spread pilots consists of transmitting low level pilot-sequences concurrently with the data. In order to reduce the cross-interferences between pilots and data, the idea is to select a pilot sequence which is orthogonal with the data sequences. This is obtained by allocating one of the WH orthogonal sequences $\textbf{c}_{p}$ to the pilots on every subset of subcarriers.
Let $\textbf{H}_{m,s}$ be the \small $\left[L\times L\right]$ \normalsize diagonal matrix of the channel coefficients associated to a given subset of subcarriers $\left[m,s\right]$. After OFDM demodulation and 2D chip de-mapping, the received signal can be expressed as:
\begin{equation}
\textbf{Z}_{m,s} = \textbf{H}_{m,s} \textbf{Y}_{m,s} + \textbf{w}_{m,s}
\end{equation}
where $\textbf{w}_{m,s} = [ w_{m,s}\left[0,0 \right] \dots w_{m,s}\left[n,q\right] \dots w_{m,s} [ L_{f}-1,L_{t}-1 ] ]^{T}$ \normalsize is the additive white Gaussian noise (AWGN) vector having zero mean and variance $\sigma^{2}_{w} = E\left\{\left|w_{m,s}\left[n,q\right]\right|^{2}\right\}$.

At the reception, the de-precoding function is processed before equalization. Therefore, an average channel coefficient $\widehat{H}_{avg}\left[m,s\right]$ is estimated by subset of subcarriers. It is obtained by de-precoding the signal received $\textbf{Z}_{m,s}$ by the pilot precoding sequence $\textbf{c}_{p}^{H}$ and then dividing by the pilot symbol $x_{m,s}^{\left(p\right)} = \sqrt{P_{p}} x_{m,s}\left[p\right]$ known by the receiver:
\begin{align} \label{H_estim}
\widehat{H}_{avg}\left[m,s\right] &= \frac{1}{x_{m,s}^{\left(p\right)}} \textbf{c}_{p}^{H} \textbf{Z}_{m,s} \nonumber \\
&= \frac{1}{x_{m,s}^{\left(p\right)}} \textbf{c}_{p}^{H} \left[\textbf{H}_{m,s} \textbf{C} \textbf{P} \textbf{x}_{m,s} + \textbf{w}_{m,s} \right] 
\end{align} \normalsize
Let us define $\textbf{C}_{u} = [\textbf{c}_{0} \dots \textbf{c}_{i\neq p} \dots \textbf{c}_{L-1} ]$ the \small $\left[L\times\left(L-1\right)\right]$ \normalsize data precoding matrix, $\textbf{P}_{u} = \text{diag} \left\{ \sqrt{P_{0}} \dots \sqrt{P_{i\neq p}} \dots \sqrt{P_{L-1}} \right\}$ \normalsize the \small $\left[ \left(L-1\right) \times \left(L-1\right) \right]$ \normalsize diagonal matrix which entries are the powers assigned to the data symbols, and \small $\textbf{x}_{m,s}^{\left(u\right)} = [ x_{m,s}[0] \dots $ \\ $x_{m,s}\left[i\neq p\right] \dots x_{m,s}\left[L-1\right] ]^{T}$ \normalsize the \small $[\left(L-1\right)\times1]$ \normalsize data symbols vector. Given these notations, (\ref{H_estim}) can be rewritten as:
%
\small
\begin{align}
\widehat{H}_{avg} & \left[m,s\right] \nonumber \\
&= \frac{1}{x_{m,s}^{\left(p\right)}} \left[ \textbf{c}_{p}^{H} \textbf{H}_{m,s} \textbf{c}_{p} x_{m,s}^{\left(p\right)} + \textbf{c}_{p}^{H} \textbf{H}_{m,s} \textbf{C}_{u} \textbf{P}_{u} \textbf{x}_{m,s}^{\left(u\right)} + \textbf{c}_{p}^{H} \textbf{w}_{m,s}  \right] \nonumber \\
&= \frac{1}{L} tr\left\{ \textbf{H}_{m,s} \right\} + \frac{1}{x_{m,s}^{\left(p\right)}} \left[ \textbf{c}_{p}^{H} \textbf{H}_{m,s} \textbf{C}_{u} \textbf{P}_{u} \textbf{x}_{m,s}^{\left(u\right)} + \textbf{c}_{p}^{H} \textbf{w}_{m,s}  \right] \nonumber \\
&= H_{avg}\left[m,s\right] + \textrm{SI}\left[m,s\right] + w'
\end{align}
\normalsize
The first term $H_{avg}\left[m,s\right]$ is the average channel response globally experienced by the subset of subcarriers $\left[m,s\right]$. The second term represents the self-interference (SI). It results from the loss of orthogonality between the precoding sequences caused by the variance of the channel coefficients over the subset of subcarriers. In the sequel, we propose to analyse its variance.

\section{THEORETICAL PERFORMANCE OF THE ESTIMATOR}
In order to analyse the theoretical performance of the proposed estimator, we evaluate its MSE under the assumption of a wide-sense stationary uncorrelated scattering (WSSUS) channel.
\begin{align} \label{MSE}
\textrm{MSE}\left[m,s\right] &= E\left\{ \left| \widehat{H}_{avg}\left[m,s\right] - H_{avg}\left[m,s\right] \right|^{2} \right\} \nonumber \\
&= E\left\{ \left| \textrm{SI}\left[m,s\right] \right|^{2} \right\} + E\left\{ \left| w' \right|^{2} \right\}
\end{align}

First, let us compute the SI variance:
\begin{equation} \label{VarSI}
E\left\{ \left| \textrm{SI}\left[m,s\right] \right|^{2} \right\} = \frac{1}{P_{p}} E\left\{ \textbf{c}_{p}^{H} \textbf{H}_{m,s} \textbf{C}_{u} \textbf{P}_{u}' \textbf{C}_{u}^{H} \textbf{H}_{m,s}^{H} \textbf{c}_{p}  \right\}
\end{equation}
where $\textbf{P}_{u}' = \textbf{P}_{u} \textbf{P}_{u}^{H} = \text{diag} \left\{ P_{0} \dots P_{i\neq p} \dots P_{L-1} \right\}$. Actually, (\ref{VarSI}) cannot be analyzed practically due to its complexity. Applying some properties of random matrix and free probability theories \cite{Tse-trans-InfoTheory} \cite{Debbah-trans-InfoTheory} which is stated in Appendix, a new SI variance formula can be derived: 
\begin{align} \label{VarSI2}
&E\left\{ \left| \textrm{SI}\left[m,s\right] \right|^{2} \right\} = \frac{1}{P_{p}} E\left\{ \textbf{c}_{p}^{H} \textbf{H}_{m,s} \left( I-\textbf{c}_{p}\textbf{c}_{p}^{H} \right) \textbf{H}_{m,s}^{H} \textbf{c}_{p} \right\} \nonumber \\
&= \frac{1}{P_{p}} E\left\{ \textbf{c}_{p}^{H} \textbf{H}_{m,s} \textbf{H}_{m,s}^{H} \textbf{c}_{p} - \textbf{c}_{p}^{H} \textbf{H}_{m,s} \textbf{c}_{p} \textbf{c}_{p}^{H} \textbf{H}_{m,s}^{H} \textbf{c}_{p} \right\} \nonumber \\
&= \frac{1}{P_{p}} E\left\{ \underbrace{ \frac{1}{L} tr\left(\textbf{H}_{m,s} \textbf{H}_{m,s}^{H}\right) }_{A} - \frac{1}{L^{2}} \underbrace{ tr\left(\textbf{H}_{m,s}\right) tr\left(\textbf{H}_{m,s}^{H}\right) }_{B} \right\} 
\end{align}
The expectation of $A$ is the average power of the channel coefficients on the subset of subcarriers $\left[m,s\right]$. Assuming that the channel coefficients are normalized, its value is one:
\begin{align} \label{A}
E\left\{ \frac{1}{L} tr\left( \textbf{H}_{m,s} \textbf{H}_{m,s}^{H} \right) \right\} &= \frac{1}{L} \sum_{n=0}^{L_{f}-1} \sum_{q=0}^{L_{t}-1 \vphantom{L_{f}}} E\left\{ \left| H_{m,s}\left[n,q\right] \right|^{2} \right\} \nonumber \\ 
&= 1
\end{align}
The expectation of $B$ is a function of the autocorrelation of the channel $R_{HH}\left(\Delta n,\Delta q\right)$ whose expression \label{Rhh_final} is developed in Appendix. Indeed, it can be written: \\
\small
\begin{equation} \label{B}
\textrm{E} \left\{ tr\left(\textbf{H}_{m,s}\right) tr\left(\textbf{H}_{m,s}^{H}\right) \right\} = \sum^{L_{f}-1}_{n=0} \sum^{L_{t}-1 \vphantom{L_{f}}}_{q=0} \sum^{L_{f}-1}_{n'=0} \sum^{L_{t}-1 \vphantom{L_{f}}}_{q'=0} R_{HH}\left(\Delta n,\Delta q\right)
\end{equation}
\normalsize
\normalsize where $\Delta n=n-n'$ and $\Delta q=q-q'$. Note that the autocorrelation function of the channel does not depend on the subset of subcarriers since the channel is WSSUS.
By combining (\ref{A}) and (\ref{B}), the SI variance expression (\ref{VarSI2}) can be expressed as:
\begin{align} \label{VarSI3}
E & \left\{ \left| \textrm{SI} \right|^{2} \right\} = \nonumber \\
& \frac{1}{P_{p}} \left( 1 - \frac{1}{L^{2}} \sum^{L_{f}-1}_{n=0} \sum^{L_{t}-1 \vphantom{L_{f}-1}}_{q=0} \sum^{L_{f}-1}_{n'=0} \sum^{L_{t}-1 \vphantom{L_{f}-1}}_{q'=0} R_{HH}\left(\Delta n,\Delta q\right) \right)
\end{align}

Now, let us compute the noise variance:
\begin{align} \label{VarNoise}
E\left\{ \left| w' \right|^{2} \right\} &= \frac{1}{P_{p}} \textrm{E} \left\{ \textbf{c}_{p}^{H} \textbf{w}_{m,s} \textbf{w}_{m,s}^{H} \textbf{c}_{p} \right\} \nonumber \\
&= \frac{1}{P_{p}} \sigma_{w}^{2}
\end{align}

Finally, by combining the expressions of the SI variance (\ref{VarSI3}) and the noise variance (\ref{VarNoise}), the MSE (\ref{MSE}) writes:
\small
\begin{align} \label{MSE2}
\textrm{MSE} = \frac{1}{P_{p}} \left(1-\frac{1}{L^{2}} \sum^{L_{f}-1}_{n=0} \sum^{L_{t}-1 \vphantom{L_{f}-1}}_{q=0} \sum^{L_{f}-1}_{n'=0} \sum^{L_{t}-1 \vphantom{L_{f}-1}}_{q'=0} R_{HH}\left(\Delta n,\Delta q\right)+\sigma_{w}^{2} \right)
\end{align}
\normalsize
The analytical expression of the MSE of our estimator depends on the pilot power, the autocorrelation function of the channel and the noise variance. The autocorrelation of the channel (\ref{Rhh_final}) is a function of both the coherence bandwidth and the coherence time. We can then expect that the proposed estimator will be all the more efficient than the channel coefficients will be highly correlated within each subset of subcarriers. One can actually check that if the channel is flat over a subset of subcarriers, then the SI (\ref{VarSI3}) is null. Therefore, it is important to optimize the time and frequency spreading lengths, $L_{t}$ and $L_{f}$, according to the transmission scenario.

\section{SIMULATION RESULTS}
In this section, we analyse the performance of the proposed 2D LP OFDM system compared to the DVB-T standard under the COST207 Typical Urban 6 paths (TU6) channel model depicted in Table \ref{TU6} with different mobile speeds. We define the parameter $\beta$ as the product between the maximum Doppler frequency $f_{D}$ and the total OFDM symbol duration $T_{\text{OFDM}}$. Table \ref{SimParam} gives the simulation parameters and the useful bit rates of the DVB-T system and the proposed system.

In the proposed system, only one spread pilot symbol is used over $L\geq16$, whereas the DVB-T system uses one pilot subcarrier over twelve. Therefore, a gain in terms of spectral efficiency and useful bit rates are obtained compared to the DVB-T system. These gains are all the higher than the spreading factor $L$ is high. Nevertheless, an increase of the spreading length produces a higher SI value. Consequently, a trade-off has to be made between the gain in term of spectral efficiency and the performance of the channel estimation.

\begin{figure}
	\begin{center}
		\includegraphics[width=1 \linewidth]{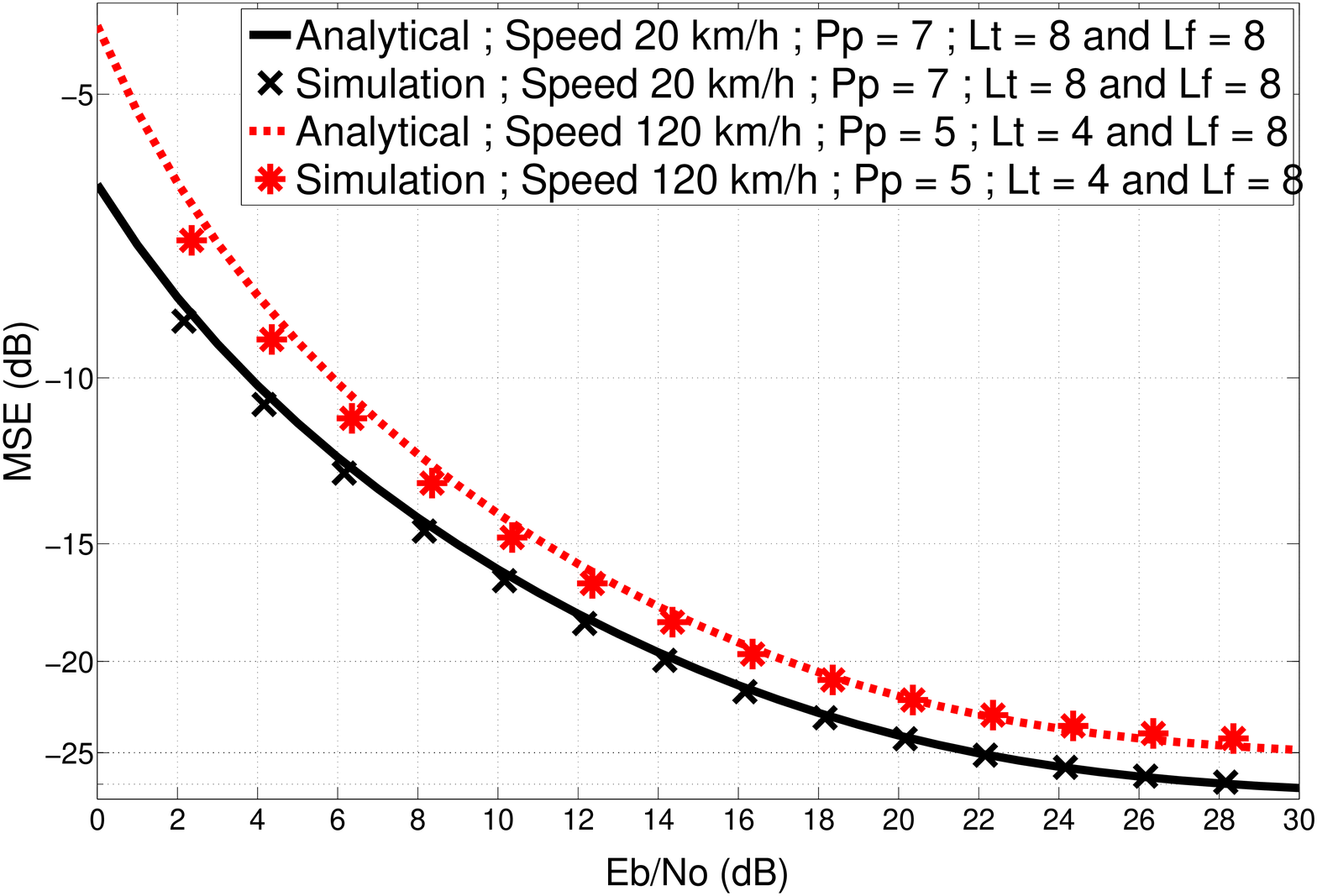}
		\caption{MSE performance obtained with the analytical expression and by simulation ; QPSK ; Speeds: 20 km/h and 120 km/h ; $\beta$ = 0.003 and 0.018}
		\label{eqm}
	\end{center}
\end{figure}

Fig. \ref{eqm} depicts the estimator performance in term of MSE for QPSK data symbols, different mobile speeds and different spreading factors. The curves represent the MSE obtained with the analytical expression (\ref{MSE2}), and the markers those obtained by simulation. We note that the MSE measured by simulation are really closed to those predicted with the MSE formula. This validates the analytical development made in section 3. We note that beyond a given ratio of the energy per bit to the noise spectral density ($\frac{Eb}{No}$), the MSE reaches a floor which is easily interpreted as being due to the SI (\ref{MSE}).

\begin{figure} [t]
	\begin{center}
		\includegraphics[width=1 \linewidth]{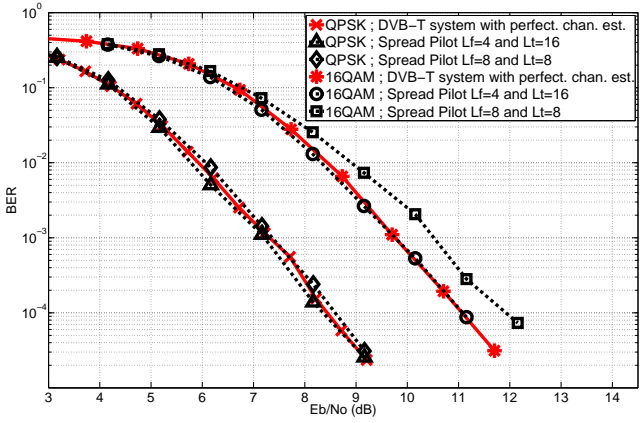}
		\caption{Performance comparison between the DVB-T system with perfect channel estimation and the proposed 2D LP OFDM ; Speed: 20 km/h ; $\beta$ = 0.003 ; $L = 64$ ; $P_{p} = 7$}
		\label{ber_speed10}
	\end{center}
\end{figure}

\begin{figure} [t]
	\begin{center}
		\includegraphics[width=1 \linewidth]{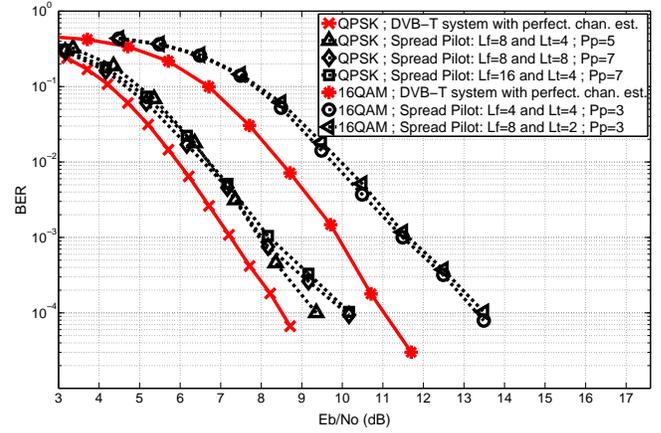}
		\caption{Performance comparison between the DVB-T system with perfect channel estimation and the proposed 2D LP OFDM ; Speed: 120 km/h ; $\beta$ = 0.018 ; $L$ = 16, 32 and 64}
		\label{ber_speed60}
	\end{center}
\end{figure}

Fig. \ref{ber_speed10} and Fig. \ref{ber_speed60} give the BER measured at the output of the Viterbi decoder for a mobile speed of 20 km/h and 120 km/h respectively. Note that the value of the pilot power $P_{p}$ has been optimized through simulation search in order to obtain the lowest BER for a given signal to noise ratio (SNR). The performance of the DVB-T system is given with perfect channel estimation, taking into account the power loss due to the amount of energy spent for the pilot subcarriers. It appears in Fig. \ref{ber_speed10}, for low-speed scenario, that the system performance is similar to that of the DVB-T system with perfect channel estimation. This is due to the power loss due to the pilot which is lower with the proposed system. In Fig. \ref{ber_speed60}, for high-speed scenario and QPSK, by choosing the spreading lengths offering the best performance, there is a loss of less than 1 dB for a $\text{BER}=10^{-4}$, comparing to perfect channel estimation case. For 16QAM, the loss is less than 2.5 dB which is really satisfying given that $\beta=0.018$, corresponding to a mobile speed of 120 km/h.


\begin{table}
\caption{Profile of TU6 channel}
\begin{center}
\begin{tabular}{|l|l|l|l|l|l|l|l|}
\hline
\small & \small Tap1 & \small Tap2 & \small Tap3 & \small Tap4 & \small Tap5 & \small Tap6 & \small unit \\
\hline
\small Delay &\small 0 &\small 0.2 &\small 0.5 &\small 1.6 &\small 2.3 &\small 5 & \small $\mu s$ \\
\hline
 \small Power &\small -3 &\small 0 &\small -5 &\small -6 &\small -8 &\small -10 & \small dB \\
\hline  
\end{tabular}
\end{center}
\label{TU6}
\end{table}
\begin{table}
\caption{Simulation Parameters and Useful Bit Rates}
\begin{center}
\begin{tabular}{|l|l|}
\hline
\small Bandwidth &\small 8 MHz \\
\hline
\small FFT size ($N_{\text{FFT}}$) &\small  2048 samples \\
\hline
\small Guard Interval size &\small  512 samples (64 $\mu$s) \\
\hline
\small OFDM symbol duration ($T_{\text{OFDM}})$ &\small  280 $\mu$s \\
\hline
\small Rate of convolutional code &\small  1/2 using $\left(133,171\right)_{o}$ \\
\hline
\small Constellations & \small  QPSK and 16QAM\\
\hline 
\small Carrier frequency &\small  500 MHz \\
\hline
\small Mobile Speeds &\small  20 km/h and 120 km/h \\
\hline
\small Maximum Doppler frequencies ($f_{D}$) &\small 9.3 Hz and 55.6 Hz \\
\hline
\small $\beta=f_{D} \times T_{\text{OFDM}}$ &\small 0.003 and 0.018 \\
\hline
\hline
\small Useful bit rates of DVB-T system 	&\small  4.98 Mbits/s for QPSK   \\
\small 												&\small  9.95 Mbits/s	for 16QAM  \\
\hline
\small Useful bit rates of 2D LP OFDM 			&\small  5.33 Mbits/s for $L=16$ \\
\small for QPSK  												&\small 	5.51 Mbits/s for $L=32$ \\
																		&\small 	5.60 Mbits/s for $L=64$ \\
\hline
\small Useful bit rates of 2D LP OFDM			&\small  10.66 Mbits/s for $L=16$ \\
\small for 16QAM  	&\small 	11.02 Mbits/s for $L=32$ \\
										&\small 	11.20 Mbits/s for $L=64$ \\
\hline
\end{tabular}
\end{center}
\label{SimParam}
\end{table}

\section{CONCLUSION}
In this paper, we propose a novel and very simple channel estimation for DVB-T. This technique, referred to as spread pilot channel estimation, allows to reduce the overhead part dedicated to channel estimation. An analytical expression of its MSE, which is a function of the autocorrelation of the channel, is given. It allows to highlight and understand that the choice of the spreading factors has to be made according to the channel characteristics. More generally, this estimation approach provides a good flexibility since it can be optimized for different mobility scenarios by choosing adequate time and frequency spreading factors. \\
\textit{This work was supported by the European project CELTIC B21C (``Broadcast for the 21st Century'').}

\par 
\bigskip
\begin{center}
\textbf{APPENDIX}
\end{center}
\par
\medskip

In this section, a property from the random matrix and free probability theories is defined for the computation of the SI variance (\ref{VarSI}). Furthermore, the computation of the autocorrelation function of the channel $R_{HH}$ is carried out.

\par
\bigskip
\noindent\textbf{Random matrix and free probability theories property}
\par
\medskip
Let $\textbf{C}$ be a Haar distributed unitary matrix \cite{Debbah-trans-InfoTheory} of size $\left[L\times L\right]$. $\textbf{C}=\left( \textbf{c}_{p},\textbf{C}_{u} \right)$ can be decomposed into a vector $\textbf{c}_{p}$ of size $\left[L\times 1\right]$ and a matrix $\textbf{C}_{u}$ of size $\left[L\times \left(L-1\right) \right]$. Given these assumptions, it is proven in \cite{Chauffray-trans-InfoTheory} that:
\begin{equation} \label{RandomMatrix}
\textbf{C}_{u} \textbf{P}_{u}' \textbf{C}_{u}^{H} \xrightarrow[]{L\rightarrow\infty} \alpha P_{u} \left( I-\textbf{c}_{p} \textbf{c}_{p}^{H} \right)
\end{equation}
where $\alpha=1$ is the system load and $P_{u}=1$ is the power of the interfering users. 

\par
\bigskip
\noindent\textbf{Autocorrelation function of the channel}
\par
\medskip
The autocorrelation function of the channel writes:
\begin{equation} \label{Rhh1}
R_{HH}\left(\Delta n,\Delta q \right) = E\left\{ H_{m,s}\left[n,q\right] H_{m,s}^{*}\left[n-\Delta n,q-\Delta q \right] \right\}
\end{equation}
We can express the frequency channel coefficients $H_{m,s}\left[n,q\right]$ as a function of the channel impulse response (CIR): 
\begin{equation} \label{FFT_CIR}
H_{m,s}\left[n,q\right] = \sum^{N_{\text{FFT}}-1}_{k=0} \gamma_{m,q}\left[k\right] e^{ -2j \pi \frac{\left(s L_{f}+n\right)}{N_{\text{FFT}}} k }
\end{equation}
where $\gamma_{m,q}\left[k\right]$ is the complex amplitude of the $k$th sample of the CIR during the $q$th OFDM symbol of the $m$th frame, and $N_{\text{FFT}}$ is the FFT size. Therefore, by injecting (\ref{FFT_CIR}) in (\ref{Rhh1}), the autocorrelation function of the channel can be rewritten as:
\small
\begin{equation} \label{Rhh2}
\begin{split}
R_{HH} & \left(\Delta n,\Delta q\right) = \\
& \frac{1}{N_{\text{FFT}}} \sum_{k=0}^{N_{\text{FFT}}-1} \sum_{k'=0}^{N_{\text{FFT}}-1} \textrm{E}\left\{ \gamma_{m,q}^{}\left[k\right] \gamma_{m,q-\Delta q}^{*}\left[k'\right] \right\} e^{-2j\pi \frac{\Delta n}{N_{\text{FFT}}} k }
\end{split}
\end{equation}
\normalsize
Since different taps of the CIR are uncorrelated, it comes:
\begin{equation} \label{Rhh3}
\begin{split}R_{HH} & \left(\Delta n,\Delta q\right) = \nonumber \\
& \frac{1}{N_{\text{FFT}}} \sum_{k=0}^{N_{\text{FFT}}-1} \textrm{E}\left\{ \gamma_{m,q}^{}\left[k\right] \gamma_{m,q-\Delta q}^{*}\left[k\right] \right\} e^{-2j\pi \frac{\Delta n}{N_{\text{FFT}}} k } 
\end{split}
\end{equation}
\normalsize
According to Jake's model \cite{Jakes}, the correlation of the $k$th sample of the CIR is: 
\begin{equation} \label{JakesEq}
\textrm{E}\left\{ \gamma_{m,q}^{}\left[k\right] \gamma_{m,q-\Delta q}^{*}\left[k\right] \right\} = \rho_{k} J_{0}\left(2\pi f_{D} \Delta q T_{\text{OFDM}} \right)
\end{equation}
where $\rho_{k}$ is the power of the $k$th sample of the CIR, $J_{0}\left(.\right)$ the zeroth-order Bessel function of the first kind, $f_{D}$ the maximum Doppler frequency and $T_{\text{OFDM}}$ the total OFDM symbol duration. 
Finally, the autocorrelation function of the channel (\ref{Rhh3}) can be expressed as: 
\begin{equation} \label{Rhh_final}
\begin{split}
R_{HH} & \left(\Delta n,\Delta q\right) = \\
& \frac{1}{N_{\text{FFT}}} \sum^{N_{\text{FFT}}-1}_{k=0} \rho_{k} e^{-2j \pi \frac{\Delta n}{N_{\text{FFT}}} k } J_{0}\left( 2\pi f_{D} \Delta q T_{\text{OFDM}} \right)
\end{split}
\end{equation}

\end{document}